\documentclass{jaa}
\usepackage{natbib}
\usepackage{xcolor}
\usepackage{graphicx}
\usepackage{hyperref}

\usepackage{etoolbox}
\apptocmd{\thebibliography}{\raggedright}{}{}

\begin{document}\sloppy

\title{Study of accretion flows around an ultraluminous X-ray source M82\,X-1 using {\it NuSTAR} data}

\author{Santanu Mondal\textsuperscript{1,*}, Biswaraj Palit\textsuperscript{1} and Sandip K. Chakrabarti\textsuperscript{2}}
\affilOne{\textsuperscript{1}Indian Institute of Astrophysics, II Block, Koramangala, Bangalore 560034, India. \\}
\affilTwo{\textsuperscript{3}Indian Centre for Space Physics, 43 Chalantika, Garia Stn. Road, Kolkata 700084, India. \\}


\twocolumn[{

\maketitle

\corres{santanuicsp@gmail.com; santanu.mondal@iiap.res.in}


\begin{abstract}
We study the spectral properties and accretion flow behavior of an ultraluminous X-ray source M82\,X-1 using {\it NuSTAR} observations. We use the physical two component advective flow (TCAF) model to fit the data and to derive the accretion flow properties of the source. From the model fitted parameters, we found that M82\,X-1 is harboring an intermediate mass black hole at its centre, where the mass varies from $156.04^{+13.51}_{-15.30}$ to $380.96^{28.38}_{-29.76}$ M$_\odot$. The error weighted average mass of the black hole is $273\pm43$ M$_\odot$, 
which accreted in nearly super-Eddington rate. The Compton cloud was compact with a size of $\sim13 r_g$ and the shock compression ratio had \textcolor{black}{nearly intermediate values except for the epoch four}. These indicate a possible significant mass outflow from the inner region of the disk. \textcolor{black}{The quasi periodic oscillation (QPO) frequencies estimated from the model fitted parameters can reproduce the observed QPOs.} The robustness of the model parameters is verified by drawing the confidence contours among them.
\end{abstract}


}]



\artcitid{\#\#\#\#}
\volnum{000}
\year{0000}
\pgrange{1--}
\setcounter{page}{1}
\lp{1}


\section{Introduction} \label{sec:intro}
Ultraluminous X-ray sources (ULXs) appear to be very luminous and non-nuclear extra-galactic objects. Most of the ULXs are thought to be X-ray binaries powered by accretion onto a central compact object which can be a stellar-mass black hole \textcolor{black}{(StMBH)} or a massive black hole (MBH), and their luminosities are comparable with or above the Eddington luminosity of stellar black holes. 
\textcolor{black}{It was also proposed that ULXs could be powered by accreting intermediate-mass black holes (IMBHs: $\sim 100-10^6$ M$_\odot$; \citet{MezcuaEtal2017} for a review) in the sub-Eddington regime \citep{Colbert1999ApJ,MatsumotoEtal2001}. 
The plausible ULX IMBH candidates might be those at the high-end of the high mass X-ray binary (HMXB) luminosity distribution \citep[see][]{MineoEtal2012MNRAS}, with luminosity $\gtrsim$ $10^{41}$ erg s$^{-1}$ in some cases.
Some of these candidates seem to show variability properties consistent with the Galactic black hole binaries \citep[GBHBs;][]{GodetEtal2009ApJ,PashamEtal2014} and the presence of cold accretion disks \citep{GodetEtal2012ApJ}, suggesting masses in the IMBH regime.
However, it is not the case always, and it has been evidenced that a large population of ULXs does not comply with canonical state transition like GBHBs \citep[][and references therein]{StobbartEtal2006MNRAS,GladstoneEtal2009}. Therefore, another possibility came up, which suggests that accretion onto \textcolor{black}{StMBH} in the super-Eddington regime could power these sources \citep{ShakuraSunyaev1973A&A,KingPound2003MNRAS,poutenan2007MNRAS.377.1187P}. Apart from extra-galactic evidence, there is now evidence for a Galactic ULX \citep{Wilson-HedgeEtal2018}. However, most of their features remain unknown.}

M82\,X-1 is an ultraluminous X-ray source present in the cigar-shaped galaxy M82, which is a star burst galaxy with a high star formation rate (SFR) $\sim$ 10M$_{\odot}$ yr$^{-1}$ \citep{kawgucgu2021A&A...652A..18I}. It is located at a distance of 3.6 Mpc \textcolor{black}{\citep{freedman1994ApJ...427..628F}. \citet{MatsumotoEtal2001} reported the presence of a low luminosity Active Galactic Nuclei (LLAGN) in M82-galaxy.} In the proximity of M82\,X-1 source lies several other X-ray luminous sources (namely X-2, X-3 and X-4 \citep[see Fig. 1;][]{BrightmanEtal2020}). M82\,X-1 \textcolor{black}{can achieve an X-ray luminosity} as high as $\sim$ 10$^{41}$ erg~s$^{-1}$, making it the brightest among all other sources. 
Most of the research on ULXs is focused on identifying the properties of the central compact object and the nature of accretion flow which is capable of producing such high X-ray luminosities. Similarly, several intriguing features of M82\,X-1 have been employed to determine its mass and accretion geometry. Analysis of Rossi X-Ray Timing Explorer (RXTE) observations by \citet{kaaret2006Sci...311..491K} had shown a 62.5 day X-ray orbital period, while \citet{Qui2015ApJ...809L..28Q} and more recently, \citet{brightmanpulsar2019ApJ...873..115B} claimed that there is degeneracy regarding the source of this orbital flux modulation and instead the period may be super-orbital in nature originating from M82\,X-2 which is now a confirmed pulsar \citep{bachetti2021arXiv211200339B}. \textcolor{black}{M82\,X-2, which is the second most luminous source in this galaxy, is only separated from M82 X-1 by a mere $5^{\prime \prime}$ angular separation.  
These two factors contribute to an enormous amount of contamination in observed properties of the source. Currently, only {\it Chandra} satellite has the resolving power required to distinguish X-1 from X-2. In order to constrain the amount of contamination, spectral studies were required to be conducted above 10 keV.}

Further, it was claimed that, M82 X-1 might be located in a young star cluster; MGG-11 \citep[]{mchardy2003ApJ...596..240M}. \citet[]{patruno2006MNRAS.370L...6P} simulated the binary evolution of this star cluster, with compact object mass range between 10-5000 M$_{\odot}$ and donor star mass range 18-26 M$_{\odot}$. The results from their simulations agreed with observations from MGG-11 and required that this dense stellar cluster hosted highly luminous BH binary systems with donor mass $\sim$ 22 M$_{\odot}$ and the BH mass range of 200-5000 M$_{\odot}$. In addition, they also argued that even with a lower IMBH range (200-400 M$_{\odot}$), it requires either beaming of disk outflow or mild super-Eddington accretion to explain its high luminosity. Hence, if M82\,X-1 belonged to this young dense cluster, then evidences \textcolor{black}{favour} an IMBH scenario.

The bolometric luminosity of M82\,X-1 exceeds the Eddington luminosity for a 10 M$_\odot$ BH by at least two orders of magnitude \textcolor{black}{\citep{dewa2006ApJ...637L..21D}}. If indeed M82\,X-1 is an IMBH accreting close to super-Eddington rate, then the mass is $\geq$ 700 M$_{\odot}$ \textcolor{black}{\citep{okajima2006ApJ...652L.105O}}. Considering a rapidly rotating BH at moderate super-Eddington accretion rates, \citet[]{fengkaaret2010ApJ...712L.169F} derived a mass range of $200 - 800$ M$_{\odot}$. Their fitting \citep[kerrbb model;][]{li2005ApJS..157..335L} results agreed with L $\propto$ T$^{4}$ variation which indicated a blackbody emission from the disk, contrasting the results of \citet[]{matsumoto2003ASPC..289..291M} who had reported no correlation among these quantities. 
However, \citet[]{okajima2006ApJ...652L.105O} pointed out that within the slim-disk formulation \textcolor{black}{(used to explain the high disk temperature)}, even a massive stellar mass BH (MStMBH) can explain spectral curvature beyond 3\,keV. Using high quality XMM-Newton observations, they reported a p-value of 0.61 (where p is the exponent in the relation T $\propto$ r$^{-p}$). Radiation hydrodynamic simulations supported the formation of disk accreting in super-Eddington rate \citep{osugha2005ApJ...628..368O}. 

A joint analysis of {\it XMM-Newton} and {\it INTEGRAL} satellite, conducted by \citet[]{SazonovEtal2014}, indicated a steep energy rollover at $\sim$ 10 keV. \textcolor{black}{This is very typical of ULXs. They also suggested the existence of cold gas along the line of site of M82\,X-1.} 
Recently, \citet{BrightmanEtal2020} carried out an extensive spectral analysis of source M82\,X-1 using simultaneous {\it Chandra}, {\it NuSTAR}, and {\it Swift/XRT} data. \textcolor{black}{Results from their spectral fitting showed that DISKPBB*SIMPL model gave the best fits to spectra of M82\,X-1 when the contamination from M82\,X-2 was less (when it was assumed to be in an off state). \textcolor{black}{In an earlier work, \citet{BrightmanEtal2016} calculated the mass }of BH in M82 X-1 from normalization-Radius relation of DISKPBB model fit \textcolor{black}{and concluded that super-Eddington accreting StMBH and estimated the mass as 26 M$_{\odot}$ and 125 M$_{\odot}$ respectively.} All estimates pointed towards a MStMBH. Overall, it was established that a puffy disk most accurately explained the spectral features of M82\,X-1, which in turn led to the conclusion that this ULX is undergoing super-Eddington accretion. Thus, two very important features of ULX have been identified, one being a high energy cutoff and other super-Eddington accretion.}

Along with the spectral variability, M82\,X-1 also showed variability in its lightcurve. Quasi-periodic oscillations (QPOs) in M82\,X-1 were first reported by \citet{stromayer2003ApJ...586L..61S} in the range of very-low frequency (LF) $\sim$ 54 mHz. Another QPO at $\sim$ 100 mHz was found by \citet[]{fiotri2004ApJ...614L.113F} associated with high/soft states ($\Gamma$ $\sim$ 2.1-2.7). Using the LFQPO frequency-Spectral Index correlation technique, they suggested that M82\,X-1 harbours intermediate mass black holes (IMBHs) $\sim$ 1000 M$_{\odot}$. \citet[]{muraciel2006MNRAS.365.1123M} showed that mHz range QPOs varied overtime, ranging from 50 mHz to 170 mHz. This corresponded to mass ranging from 10 M$_{\odot}$ to 1000 M$_{\odot}$. \textcolor{black}{The mass estimation of 25-520 M$_{\odot}$ done by \citet{dewa2006ApJ...637L..21D} using the photon index v/s QPO frequency relation falls in the above range.} More recently, ``twin-peaked" stable 3:2 ratio QPOs; 3.32 Hz and 5.07 Hz were detected for this source by \citet[]{PashamEtal2014}; yielded mass roughly $\sim$ 400 M$_{\odot}$.

Since the proposition of mass outflow from inner regions of an advected disk in ULXs by \citet[]{poutenan2007MNRAS.377.1187P}, several ULXs (NGC 1313 X-1 \textcolor{black}{\citet{pinto2016Natur.533...64P}}, NGC 5408 X-1 \textcolor{black}{\citet{pinto2016Natur.533...64P}}, and NGC 6946 X-1 \textcolor{black}{\citet{Kosec2018MNRAS.473.5680K}}) have been found to exhibit soft-excess below 2 keV, which indicates a wind dominated accretion disk. However, since the soft X-ray emission from M82\,X-1 coincides with the diffuse plasma emission \citep[]{miller22005Ap&SS.300..227M}, spectral information below 2 keV cannot be trusted. Hence, it has been difficult to study M82 X-1 in the context of wind launching from accretion disks. Only recently \citet[]{BrightmanEtal2020} has hinted at a possible positive correlation between mass outflow rate and absorption by galactic column density.
So far, we noticed that investigating temporal properties have mostly favoured IMBH while simple multicoloured DISK+POWERLAW continuum models or hybrid-disk models DISKPBB have favoured \textcolor{black}{MStMBHs}. The accretion onto these objects is exhaustively studied, and several phenomenological models \citep[see][for a review]{KaaretEtal2017ARA&A} are proposed in the literature, however, we know very little about the disk formation, origin of inner hot region, its optical depth, geometry, etc. Therefore, we require a self-consistent models that are able to explain the harder energy range spectra, which are less contaminated. 

Despite the fact that the source has been studied in different energy bands, we know very little about the underlying physical processes that created such variable spectral features. Also, the literature estimated a broad range of mass of the compact object, leaving a range of questions: (1) does this system harbour a StMBH or an IMBH? (2) what is the nature of accretion behind such a high luminosity? (3) is the high column density along the line of sight (LOS) due to the wind from the disk, or due to some moving cloud along the LOS? In this paper, we aim to address different observational aspects after performing spectral fitting of {\it NuSTAR } data using two component advective flow (TCAF) model \citep{chak1995ApJ...455..623C}. The paper is organized as follows: In Section 2, we discuss the observation log, data analysis procedures, \textcolor{black}{and modelling}. In Section 3, we present our spectral fitting results and different estimations. Finally, in Section 4, we make our concluding remarks.

\section{Observation and data analysis}
For the spectral analysis \textcolor{black}{of M82\,X-1 (located at RA=148.958 and Dec=+69.679 in the galaxy M82)} of each epoch of observation, we used the data of {\it NuSTAR} \citep{Harrisonetal2013} during 2016 in the energy range of 3$-$30\,keV.
{\it NuSTAR} data was extracted using the standard \textcolor{black}{NuSTAR Data Analysis Software} ({\sc NuSTARDAS v1.3.1})
\footnote{\url{https://heasarc.gsfc.nasa.gov/docs/nustar/analysis/}} software. \textcolor{black}{The exposure time and the observation log are given in \autoref{tab:ObsLog}.} We ran {\sc nupipeline} task to produce cleaned event lists and 
{\sc nuproducts} \textcolor{black}{task} to generate \textcolor{black}{the source and background spectra with the standard filters. Source events were selected from a circular region of 50$^{\prime \prime}$. The background region were selected from larger circular source-free region (of 60$^{\prime \prime}$) to avoid contamination.} The data was grouped by {\it grppha} command, with a 
minimum of 25 counts in each bin. To avoid contamination of the source from X-2, we considered those observations which have much higher flux compared to X-2, following the data selection discussed in \citet{BrightmanEtal2020}.

We used {\sc XSPEC}\footnote{\url{https://heasarc.gsfc.nasa.gov/xanadu/xspec/}} \citep{Arnaud1996} version 12.11.0 for spectral analysis. 
Each epoch of observation was fitted using both phenomenological and physical models, which is discussed in the later sections. \textcolor{black}{For all observations, we used the absorption model 
tbabs \citep{Wilmsetal2000}, for the Galactic absorption,} keeping the hydrogen column density ($N_{H}$) fixed to $0.9\times 10^{22}$cm$^{-2}$ \textcolor{black}{\citep[][the Leiden/Argentine/Bonn (LAB) survey]{BrightmanEtal2020,Kalberlaetal2005}} during the fitting. \textcolor{black}{In addition to the above $N_{H}$, the host galaxy $N_{H}$ is also taken into account using the partial covering ionization (ZXIPCF) model \textcolor{black}{\citep{ReevesEtal2008}}, which is discussed later.} We used chi-square statistics for the goodness of the fitting.

\begin{table}
\scriptsize
    \centering
     \caption{{\it NuSTAR} observations used in this paper}
    \begin{tabular}{c c c c c}
    \hline
           Epoch & OBSID & Date & MJD & Exposure\\
           \hline
 E1 & 80202020002 & 2016-01-26 & 57413 & 36 ksec\\
 E2 & 30202022002 & 2016-06-03 & 57542 & 39 ksec\\
 E3 & 30202022008 & 2016-07-29 & 57598 & 42 ksec\\
 E4 & 90202038002 & 2016-10-07 & 57688 & 45 ksec\\
      \hline
    \end{tabular}
      \label{tab:ObsLog}
\end{table}

\subsection{Modelling with TCAF}
\textcolor{black}{There are several models available in the literature which have been used to study the source M82\,X-1. Alternatively, a more physical model, namely a TCAF model is also present in the literature and this provides a natural explanation of two phase configuration.} For the spectral fitting, we used TCAF model along with a
multiplicative ZXIPCF model.

According to TCAF, matter falling onto a BH has two components, one is the high angular momentum Keplerian flow sitting at the equatorial plane and the second component is the hot, low angular momentum, sub-Keplerian flow, which forms the inner hot region after forming a shock due after satisfying Rankine-Hugoniot shock conditions \citep[see][]{Chakrabarti1989}. The boundary layer of this region has physical origin rather than phenomenological, which is called CENtrifugal pressure supported BOundary Layer (CENBOL). This model not only considers accretion, but also launches jet in a self-consistent way; therefore it can explain accretion-ejection or disk-jet connection in BH systems \citep[see][]{Chakrabarti1999,MondalChakrabarti2021}.

The model requires five parameters (if the mass is unknown and is considered to be a free parameter) which are (i) mass of the BH ($M_{\rm BH}$), (ii) disk accretion rate 
($\dot m_{\rm d}$), (iii) halo accretion rate ($\dot m_{\rm h}$), (iv) size of the CENBOL or the location of the shock ($X_{\rm s}$ in $r_g=2GM_{\rm BH}/c^2$ unit), and (v) shock compression ratio ($R$). In this model two-temperature equations for the electron and protons are solved to compute the inverse-Comptonized spectra. The emergent spectrum from TCAF has mainly three components, (1) multicolour blackbody spectrum which is coming from the disc, (2) the hard radiation; from the upscattering of the soft photons from the disk by the hot corona, and (3) the scattering of the hard radiation by the cold disc; which is so-called reflection component. Till date, TCAF model has been successfully applied to study the outburst properties of StMBHs \citep[][and references therein]{Debnathetal2014,Mondaletal2014,Janaetal2016}, AGNs \citep{Nandietal2019,MondalStalin2021}, and estimated the mass of the central BHs \citep{molla2017ApJ...834...88M,Nandietal2019}. In the timing domain, the oscillation of the shock in this model self-consistently explains the occurrence of QPOs, and this interpretation in the context of thick disks was proposed by \citet[][]{molteni1996ApJ...457..805M}. \textcolor{black}{Later, the shock oscillation model was verified and applied extensively to several GBHBs
\citep[][and references therein]{chakraborti2015MNRAS.452.3451C,Mondal2020MNRAS,chandra2021}. Considering the success of this model in explaining the observed spectral and temporal properties of GBHBs, we now employ it to the ULX M82\,X-1 as  ULXs have been compared with GBHBs.} M82\,X-1 also showed two Type-C QPO frequency, $\sim$ 3.32 Hz and 5.07 Hz \citep[]{PashamEtal2014}. All these observational findings and evidence have prompted us to use this model and study the nature of accretion flow of M82 X-1 which gives rise to such a high luminosity.

\section{Results}
\subsection{Spectral fit}
\begin{figure*}[t!]
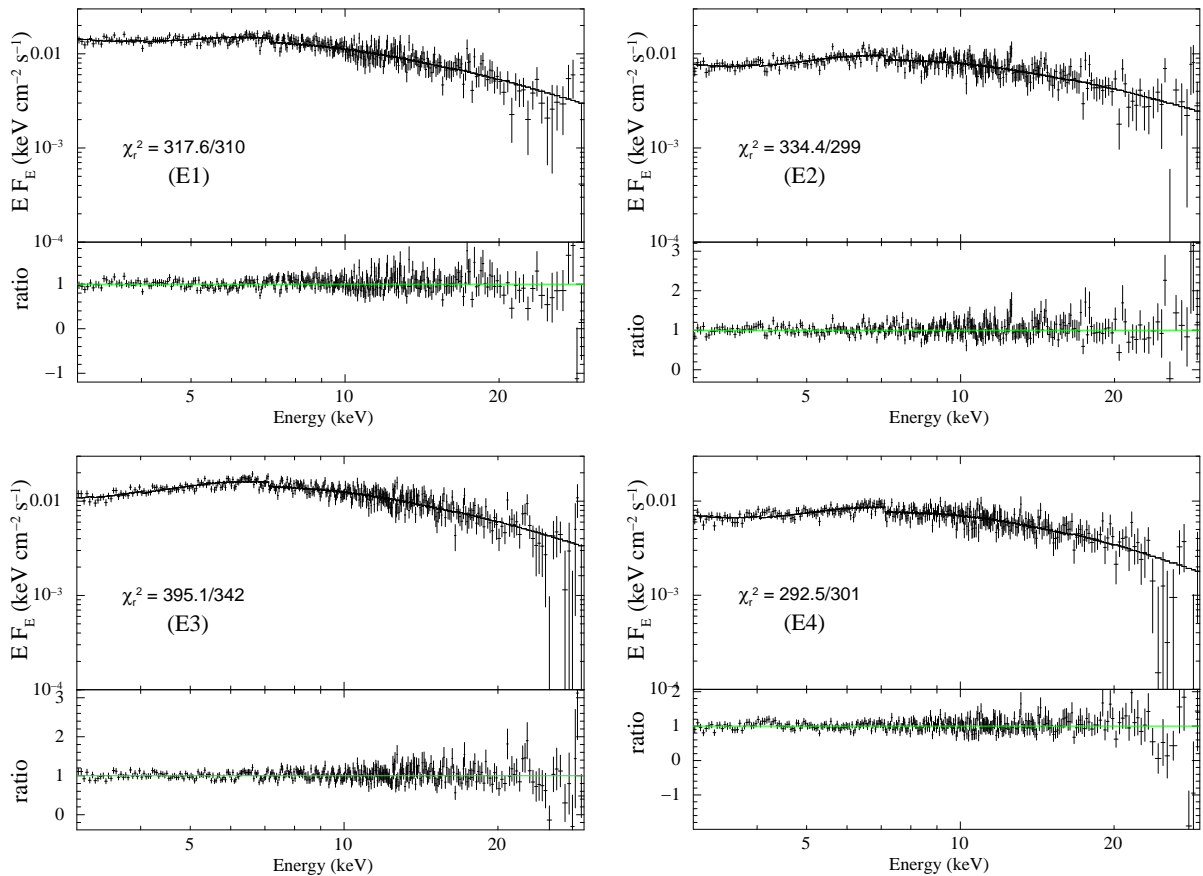

\centering{ 
\includegraphics[height=8truecm,angle=270]{802-tcaf-zxipcf-2.eps}
\includegraphics[height=8truecm,angle=270]{30202-tcaf-zxipcf-2.eps}} 
\centering{ 
\includegraphics[height=8truecm,angle=270]{308-tcaf-zxipcf-2.eps}
\includegraphics[height=8truecm,angle=270]{902-tcaf-zxipcf-2.eps}}
\caption {TCAF model fitted 3$-$30.0 keV spectra for all four epochs of M82\,X-1 during {\it NuSTAR} era. }
\label{fig:SpecTcaf}
\end{figure*}

\begin{figure}
\hspace{-2.2cm}
    \centering
    \includegraphics[height=10.0cm]{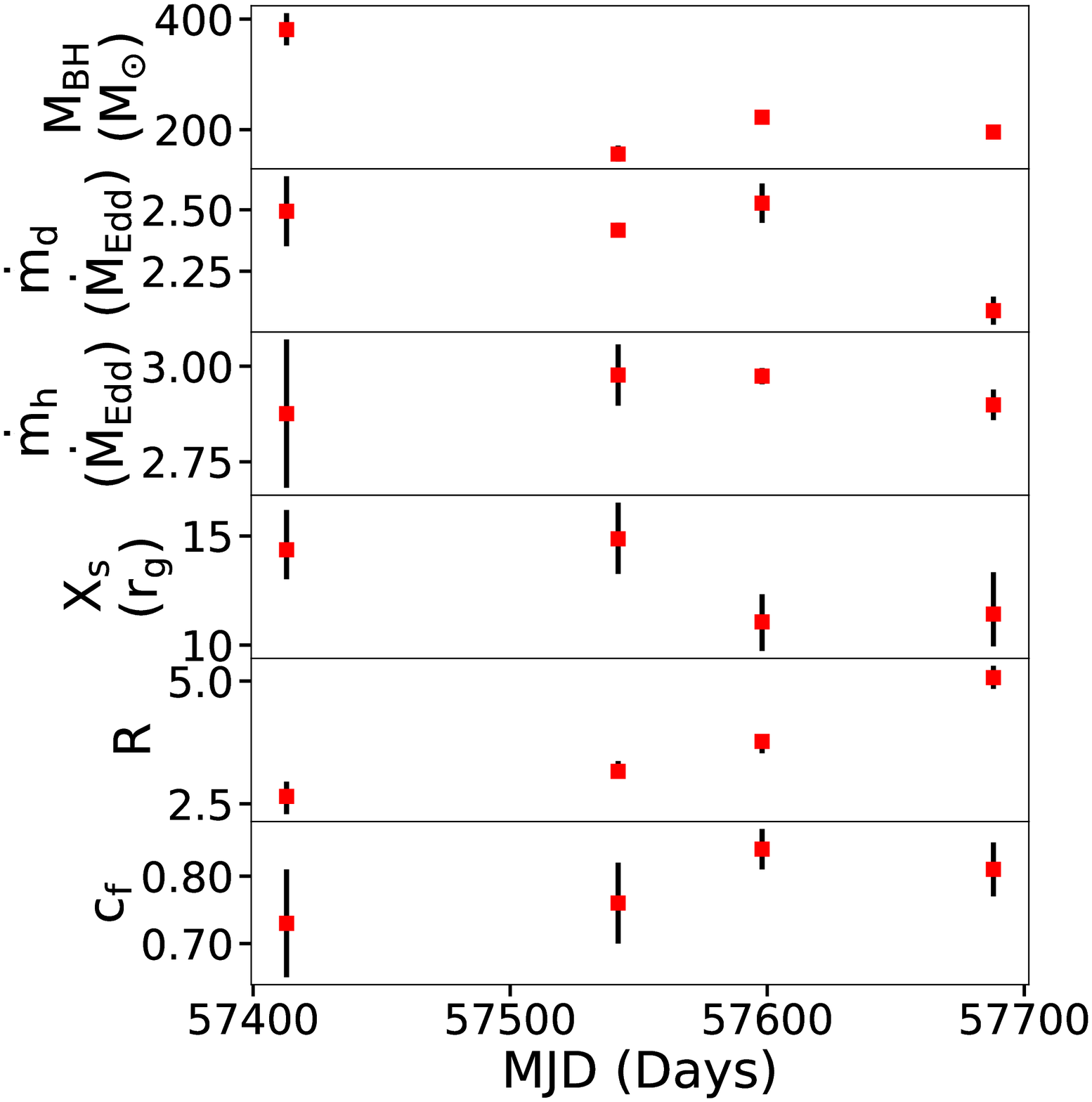}
    \caption{Variation of the best fitted TCAF model parameters with MJD are shown. The bottom panel shows the partial covering fraction parameter of ZXIPCF model.}
    \label{fig:TcafPars}
\end{figure}

\textcolor{black}{In \autoref{fig:SpecTcaf}, we show the TCAF model fitted 3-30 keV spectra for all four epochs. We have obtained acceptable fit statistics with reduced $\chi^{2}$ values ranging from $0.97$ to $1.15$, mentioned with the figures. The data quality is noisy above 25 keV. The corresponding model fitted parameters are provided in \autoref{tab:TCAFResults}.}
It can be seen that the mass of the BH varied in a range between 156 M$_\odot$ and 381 M$_\odot$, the error weighed value is $273\pm43M_\odot$. 
This implies that M82 is harboring an IMBH at its centre. The mass accretion rates we obtained are in the super-Eddington regime \textcolor{black}{with values of $\dot m_{\rm d}$ from 2.09 to 2.53 $\dot M_{\rm Edd}$ and $\dot m_{\rm h}$ from 2.88 to 2.98 $\dot M_{\rm Edd}$}. This agrees with the previous claims of M82\,X-1 being a super-Eddington accretor. The CENBOL was compact, varies in a narrow range between $11$ and 15 $r_g$. The intermediate values of $R$, indicate a possible mass outflow from the system \citep{Chakrabarti1999}. In \autoref{fig:TcafPars}, variation of all model parameters with MJD are shown. \textcolor{black}{The top five panels show the variation of the mass of the BH, disk and halo accretion rates, location of the shock and shock compression ratio respectively, which are variable during the observation period. The bottom panel shows covering fraction from ZXIPCF model.} \textcolor{black}{Considering epochs E3 and E4, which are separated by 68 days, we notice a significant change in $\dot m_{\rm d}$ from 2.42 to 2.09 $\dot M_{\rm Edd}$, and $R$ from 3.77 to 5.07. However, $X_s$ remained more or less the same, which can be due to decrease in $\dot m_{\rm h}$. On the contrary, in case of epochs E2 and E3, separated by 56 days, increase in $\dot m_{\rm d}$ from 2.42 to 2.53 $\dot M_{\rm Edd}$} shifts the shock location towards the BH, \textcolor{black}{when the $\dot m_{\rm h}$ remained the same.} 

It should be noted that the ZXIPCF model fitted N$_H$ along the LOS and covering factor ($C_f$) are significantly \textcolor{black}{variable and} high. The super-Eddington mass accretion naturally explains that the wind was launched from the disk, which blocked the central continuum radiation, therefore N$_H$ is also high \textcolor{black}{(col. 8 in \autoref{tab:TCAFResults}).} 
\textcolor{black}{The anti-correlation between N$_H$ and $\dot m_{\rm d}$ can explain the continuum luminosity driven outflow from the disk. The increase in $\dot m_{\rm d}$, increases the luminosity of the continuum. Assuming that the photon momentum is transferred to the wind from the disk, therefore it increases the momentum of the wind outflow as well.} A higher momentum flow can escape the gravitational barrier of the disk and give a low value of N$_H$ along the LOS, whereas, the effect is opposite when $\dot m_{\rm d}$ is low \citep[see an active galactic nucleus case study in][]{MondalEtAl2022CLAGN}.  However, unlike many other ULXs, it has not been able to study M82\,X-1 in the context of wind launching from the inner disk from absorption line variability \citep[][]{pinto2016Natur.533...64P}, mainly due to diffuse emission of the background.

\begin{table*}
\centering
\caption{Best fitted TCAF model parameters are provided. Here, N$_H$ and C$_f$ are hydrogen column density and dimensionless covering fraction for the partial covering model respectively.}
\resizebox{\textwidth}{!}{\begin{tabular}{cccccccccc}
\hline
Obs Ids. &\textcolor{black}{Epoch}    &$M_{\rm BH}$ &$\dot m_{\rm d}$ & $\dot m_{\rm h}$ & $X_{\rm s}$ & R &$N_H$&C$_f$&$\chi^2/dof$ \\
	     &   & $M_\odot$&$\dot M_{\rm Edd}$&$\dot M_{\rm Edd}$&$r_{\rm g}$& &$\times 10^{22}$ cm$^{-2}$& & \\
\hline
80202020002&E1 &$380.96^{+28.38}_{-29.76}$&$2.494^{+0.142}_{-0.102}$&$2.876^{+0.194}_{-0.192}$&$14.37^{+1.35}_{-1.82}$&$2.65^{+0.36}_{-0.30}$&$21.29^{+1.16}_{-1.38}$&$0.73^{+0.08}_{-0.05}$&317.6/310\\
30202022002&E2 &$156.04^{+13.51}_{-15.30}$&$2.417^{+0.03}_{-0.03}$&$2.977^{+0.08}_{-0.07}$&$14.87^{+1.61}_{-1.66}$&$3.16^{+0.13}_{-0.21}$&$23.77^{+1.13}_{-1.34}$&$0.76^{+0.06}_{-0.10}$&334.4/299 \\
30202022008&E3 &$222.57^{+10.76}_{-13.14}$&$2.527^{+0.08}_{-0.07}$&$2.974^{+0.021}_{-0.106}$&$11.06^{+1.33}_{-1.27}$&$3.77^{+0.24}_{-0.12}$&$18.23^{+0.91}_{-1.12}$&$0.84^{+0.03}_{-0.04}$&395.1/342 \\
90202038002&E4 &$195.75^{+4.91}_{-4.75}$&$2.090^{+0.057}_{-0.056}$&$2.899^{+0.041}_{-0.042}$&$11.42^{+1.48}_{-1.92}$&$5.07^{+0.23}_{-0.24}$&$28.47^{+0.61}_{-0.59}$&$0.81^{+0.04}_{-0.04}$&292.5/301 \\
\hline
\end{tabular}\label{tab:TCAFResults}  }
\end{table*}

\begin{figure}
    \centering
    \includegraphics[height=12.0cm]{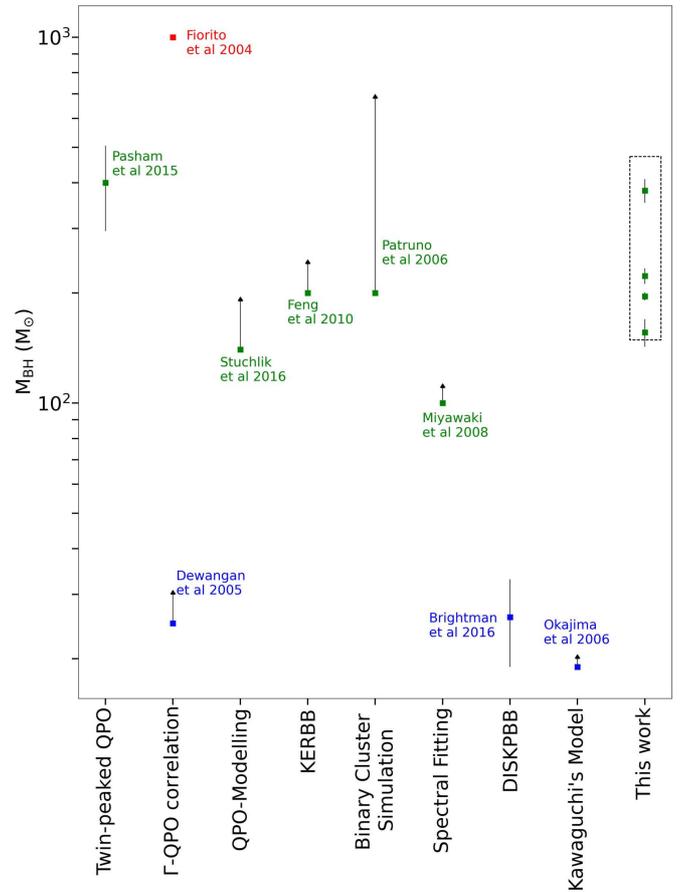}
    \caption{All possible estimates of the mass of M82\,X-1 obtained from different analyses in the literature. Red, green, and blue colored data points represent mass estimates above 1000M$_\odot$, between 1000 M$_\odot$-100 M$_\odot$ and less than 100 M$_\odot$ respectively. Authors who have only reported a range of M$_{\rm BH}$s, for them we have only plotted the lower limit and extended an arrow upwards towards the upper limit where the length of the arrow is proportional to range of the mass estimation. X-axis indicates the models used by the authors. \textcolor{black}{The points inside the rectangle represent the mass estimated in this paper.}}
    \label{fig:massplot}
\end{figure}

\textcolor{black}{The BH mass proposed for M82\,X-1 varies in a significantly broad range between $\sim $10 to $10^3$ M$_\odot$. However, TCAF model fitted BH mass varies in a narrow range.} To compare the TCAF model fitted BH mass with that present in the literature, we plot all possible estimated masses in \autoref{fig:massplot}. \textcolor{black}{As discussed in \autoref{sec:intro}, the powering of ULXs can be explained depending on the accretor mass and its mass accretion rates at different regimes. In addition to other possibilities, our spectral study shows that an IMBH can also accrete in the super-Eddington limit. This estimate} agrees with the claims in the literature that super-Eddington accretion may not be limited to stellar mass BHs \citep{kaaret2006ApJ...646..174K}. It is evident that super-massive BHs can accrete in super-Eddington limit \citep[][and references therein]{PuDu2015ApJ...806...22D, hlieu2021ApJ...910..103L} and the physical processes in low mass BHs to the supermassive BHs can be scaled by their mass \citep{McHardyetal2006}. Furthermore, in principle, advective flows can accrete in the super-Eddington regime. Observational evidence of the IMBH AGN source-RX J1140.1+0307 suggests it is accreting at $\sim$ 10 times its Eddington limit \citep{jin2016MNRAS.455..691J}. One of the often discussed models for growth of BHs into \textcolor{black}{super-massive BHs (SMBHs)} is through rapid mass accretion in $\sim$ 100 M$_\odot $ BHs \citep{jenny2020ARA&A..58..257G, Toyouchi2021ApJ...907...74T}. Such IMBHs may be formed after the death of the earliest known Pop-III stars. Therefore, super-Eddington accretion onto IMBHs is a necessary requirement to form SMBHs within appropriate timescales \citep[][and references therein]{jenny2020ARA&A..58..257G}. \textcolor{black}{To further strengthen our claim we have drawn confindence contour of the models parameters in the next section.}

\subsection{Robustness of model fitted parameters}
\begin{figure*}[t!]
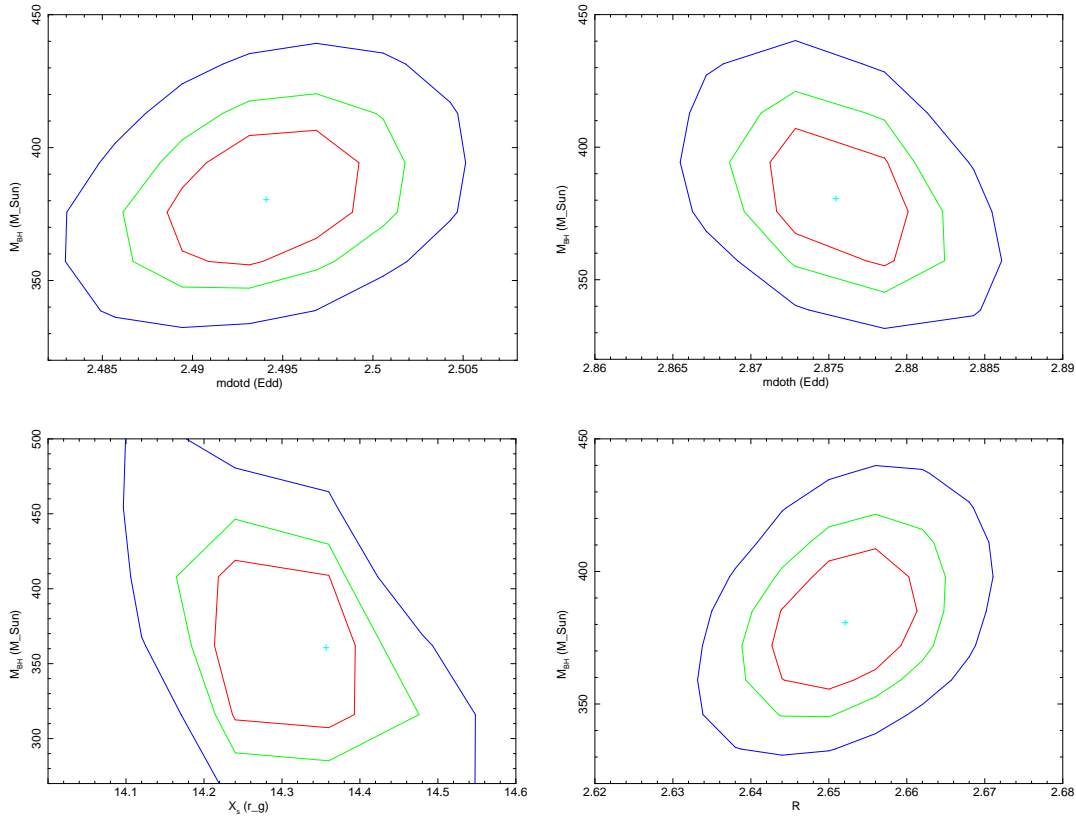

\centering{ 
\includegraphics[height=8truecm,angle=270]{802-mbh-md-contour.eps}
\hspace{-1.0cm}
\includegraphics[height=8truecm,angle=270]{802-mbh-mh-contour.eps}} \\
\centering{ 
\includegraphics[height=8truecm,angle=270]{802-mbh-xs-contour.eps}
\hspace{-1.0cm}
\includegraphics[height=8truecm,angle=270]{802-mbh-R-contour.eps}}
\caption {Confidence contours of $M_{\rm BH}$ with other model parameters for the epoch E1 are shown. Three different contour colors show the confidence level of one (red), two (green), and three sigma (blue). Here, x-labels follow the parameter notations used in \autoref{tab:TCAFResults}, e.g. mdotd, mdoth, and $Xs$ represents $\dot m_d$, $\dot m_h$, and $X_s$ respectively. }
\label{fig:ContourTcaf}
\end{figure*}

\autoref{fig:ContourTcaf} shows the confidence contours of $M_{BH}$ with other TCAF model parameters. We have used {\it steppar} task in {\it XSPEC} to generate these plots. \textcolor{black}{The top panels show} the confidence contours of $M_{BH}$ with disk and halo accretion rates, and \textcolor{black}{the bottom panels show} the correlation with the location of the shock and its compression ratio. Different model parameters are labelled in x-axis. Three different contour colors (red, green, and blue) correspond to one, two, and three sigma confidence levels. It is also the same as $\Delta \chi^2$ fit statistics of 2.3, 4.61, and 9.21. The correlation of $M_{BH}$ with other model parameters are clearly visible from the contour plots. The model parameters assess better precision. The correlations are clear, which show that there is no obvious degeneracy between parameters and the solution is robust. \textcolor{black}{This additional test reasonably agrees our new possibility of powering the ULXs.}

\subsection{QPO frequency from model parameters}
As explained by the propagatory shock oscillation model \citep{ChakrabartiEtal.POS2005,chakrabartiEtal2008A&A...489L..41C}, 
the QPO frequencies follow the relation below: 
\begin{equation}
   \nu_{\rm QPO}= \frac{c}{r_{\rm g}\; M_{\rm BH}\; R X_{\rm s}^{1.5}}
\end{equation}
Considering model fitted parameters from \autoref{tab:TCAFResults}, we have estimated the QPO frequencies for all four epochs, the $\nu_{\rm QPO}$ values range from 1.83 Hz to 3.53 Hz (see, \autoref{tab:QPOmass}). The error weighed $\nu_{\rm QPO}$ is 2.93 $\pm$ 0.32 Hz. In epochs E2 and E3, the values are closer to the lower twin-peaked QPO (3.32 Hz) estimated by \citet[][]{PashamEtal2014} from power density spectra. Therefore, our model fitted parameters are also able to reproduce the observed QPOs. Conversely, it can be also be inferred that \textcolor{black}{BH} mass estimated from the QPO frequency is along the line of our model fitted mass. As evident from \autoref{fig:massplot}, mass values of M82\,X-1 dominate the 100M$_\odot$ - 400M$_\odot$ region, consistent with our estimate.

\begin{table}
\large
    \centering
    \caption{Estimated QPO frequencies against those reported by \citet[][]{PashamEtal2014}}
    \begin{tabular}{c c c}
    \hline
    Epoch & $\nu_{\rm QPO}$ & $\nu_{\rm QPO}$ \\
    & Estimated &Observed   \\  
    &Hz &Hz\\
    \hline
    E1  &1.83{\raisebox{0.5ex}{\tiny$\substack{+0.38 \\ -0.42}$}} & \\
    E2 &3.53{\raisebox{0.5ex}{\tiny$\substack{+0.66 \\ -0.72}$}} &3.32  {\raisebox{0.5ex}{\tiny$\substack{+0.06 \\ -0.06}$}}\\
    E3 &3.24 {\raisebox{0.5ex}{\tiny$\substack{+0.63 \\ -0.59}$}} &5.07 {\raisebox{0.5ex}{\tiny$\substack{+0.06 \\ -0.06}$}}\\
    E4 &2.61 {\raisebox{0.5ex}{\tiny$\substack{+0.52 \\ -0.67}$}} &\\
   \hline
    \end{tabular}
    \label{tab:QPOmass}
\end{table}

\section{Conclusions}
In this paper we study the accretion properties of the ultraluminous X-ray source M82\,X-1 using {\it NuSTAR} data with the TCAF model. Our main conclusions are follows:

\begin{itemize}
    \item TCAF model can successfully fit and explain the ULXs data in the same way as it did for Galactic black hole binaries and active galactic nuclei.
    
    \item The model fitted mass falls in the intermediate BH mass range between 156 M$_\odot$ and 381 M$_\odot$. 
    
    \item Our estimates suggest that both the Keplerian and sub-Keplerian mass accretion rates are nearly in the super-Eddington range.
    
    \item The CENBOL was found to be compact around ~13 r$_g$ during the observation period.
    
    \item The shock compression ratio ($R$) falls in the intermediate range indicates that the mass outflow from the CENBOL region was significant \citep{Chakrabarti1999}.
    
    \item The confidence contours of mass with other parameters points to the robustness of the estimation of the mass and the other model parameters.
    
    \item The super-Eddington mass accretion can be the reason of disk wind as observed in the literature.
    
    \item \textcolor{black}{The QPO frequencies estimated from the TCAF model fitted parameters agree well with the observations in the literature.}
    
\end{itemize}

\textcolor{black}{This study simultaneously explains both spectral and temporal properties of the ULX M82\,X-1 using the same model parameters which have been obtained from spectral fitting.}

\section*{Acknowledgements}
\textcolor{black}{We thank the referee for making constructive comments and suggestions. SM thanks M. Brightman for discussions on resolving the source.} SM and BP acknowledge Ramanujan Fellowship research grant (File \#RJF/2020/000113) by DST-SERB, Govt. of India for this research. This research has made use of the {\it NuSTAR} Data Analysis Software ({\sc nustardas}) jointly developed by the ASI Science Data Center (ASDC), Italy and the California Institute of Technology (Caltech), USA. This research has also made use of data obtained through the High Energy Astrophysics Science Archive Research Center Online Service, provided by NASA/Goddard Space Flight Center.

%



\bibliography{m82x1} 








\end{document}